\documentclass[useAMS,usenatbib]{mn2e}

\usepackage{color}
\usepackage{graphicx}
\usepackage{graphics}
\usepackage{rotating}
\usepackage{amsmath}        
\usepackage{amssymb}        
\usepackage{subfigure}        
\usepackage{epsfig}        

\newcommand{\degree}{$^{\circ}$}
\newcommand{\arcminute}{$^{\prime}$}
\newcommand{\arcsecond}{$^{\prime\prime}$}
\newcommand{\perbeam}{\,beam$^{-1}$}

\title[Astrometry of XTE J1752-223]{An accurate position for the black hole candidate XTE J1752-223: re-interpretation of the VLBI data}

\author[J.C.A.~Miller-Jones et al.]
 {J.C.A.~Miller-Jones,$^{1}$\thanks{email: james.miller-jones@curtin.edu.au}
 P.G.~Jonker,$^{2,3,4}$ E.M.~Ratti,$^2$ M.A.P.~Torres,$^{2,3}$ C.~Brocksopp,$^{5}$ \and J.~Yang,$^{6}$ and N.I.~Morrell$^{7}$\\
$^1$International Centre for Radio Astronomy Research - Curtin University, GPO Box U1987, Perth, WA 6845, Australia\\
$^2$SRON, Netherlands Institute for Space Research, 3584 CA,
  Utrecht, the Netherlands\\
$^3$Harvard-Smithsonian Center for Astrophysics, 60 Garden Street, Cambridge, MA 02138, USA\\
$^4$Department of Astrophysics, IMAPP, Radboud University Nijmegen,
Heyendaalseweg 135, 6525 AJ, Nijmegen, The Netherlands\\
$^5$Mullard Space Science Laboratory, University College London, Holmbury St. Mary, Dorking, Surrey RH5 6NT, UK\\
$^6$Joint Institute for VLBI in Europe, Postbus 2, 7990 AA, Dwingeloo, The Netherlands\\
$^7$Las Campanas Observatory, The Carnegie Observatories, Casilla 601, La Serena, Chile\\
}
\begin{document}

\date{Accepted 2011 March 12.  Received 2011 March 9; in original form 2011 January 23}

\pagerange{\pageref{firstpage}--\pageref{lastpage}} \pubyear{2011}

\maketitle

\label{firstpage}

\begin{abstract}
Using high-precision astrometric optical observations from the Walter Baade Magellan Telescope in conjunction with high-resolution very long baseline interferometric (VLBI) radio imaging with the Very Long Baseline Array (VLBA), we have located the core of the X-ray binary system XTE J1752-223.  Compact radio emission from the core was detected following the state transition from the soft to the hard X-ray state.  Its position to the south-east of all previously-detected jet components mandated a re-analysis of the existing VLBI data.  Our analysis suggests that the outburst comprised at least two ejection events prior to 2010 February 26.  No radio-emitting components were detected to the south-east of the core at any epoch, suggesting that the receding jets were Doppler-deboosted below our sensitivity limit.  From the ratio of the brightness of the detected components to the measured upper limits for the receding ejecta, we constrain the jet speed $\beta>0.66$ and the inclination angle to the line of sight $\theta<49^{\circ}$.  Assuming that the initial ejection event occurred at the transition from the hard intermediate state to the soft intermediate state, an initial period of ballistic motion followed by a Sedov phase (i.e.\ self-similar adiabatic expansion) appears to fit the motion of the ejecta better than a uniform deceleration model.  The accurate core location can provide a long time baseline for a future proper motion determination should the system show a second outburst, providing insights into the formation mechanism of the compact object.
\end{abstract}


\section{Introduction}

Accurate astrometry is crucial in interpreting astronomical observations, allowing the association of sources observed in one region of the electromagnetic spectrum with their counterparts at other wavebands, and thus providing the full spectral energy distribution of an astrophysical object.  For Galactic objects such as X-ray binaries, high-precision astrometry can also provide estimates of the proper motion of the source \citep[e.g.][]{Mir01,Mir02,Mir03,Dha07,Mil09a}, and even the source distance via trigonometric parallax \citep[e.g.][]{Bra99,Mil09b}.  At present, the highest precision can be achieved in the radio band using the technique of very long baseline interferometry (VLBI).  However, the high resolution of VLBI arrays that enables such high-precision astrometry also makes it impractical to use this technique to search for sources without a well-defined error circle from lower-resolution instruments, since the number of pixels to be searched can become unfeasibly large.

Black hole X-ray binaries undergo occasional outbursts in which relativistically-moving jet knots are seen to move away from the central binary system.  Accurate astrometry is also required to interpret observations of these jets, since knowledge of the position of the central binary allows us to compare the relative motion of approaching and receding jet components and hence to constrain the product $\beta\cos\theta$, where $\beta$ is the jet speed as a fraction of the speed of light, and $\theta$ is the inclination angle of the jet axis to the line of sight \citep{Mir94}.  Knowledge of the Doppler factor in turn helps constrain the energetics of the jets.

As they evolve through their duty cycles, black hole X-ray binaries pass through a range of canonical states, defined by their X-ray spectral and timing characteristics.  These X-ray states, representative of specific conditions within the accretion flow, are very well correlated with the behaviour of the associated outflow in the form of radio jets \citep*[for a detailed discussion of the disc-jet coupling, see][]{Fen04}.  In the low/hard X-ray spectral state, observed at the beginning and end of an outburst, a steady, compact, flat-spectrum jet exists, which subsequently gives rise to bright, relativistically-moving, discrete ejecta as the X-ray spectrum softens at the peak of the outburst and the source moves through hard and soft intermediate states into a high/soft state.  The compact radio jet is then quenched, with no core radio emission being detected until the source moves back from the high/soft state through the intermediate states into the low/hard state once more.  The quenching of the core radio emission at the peak of the outburst implies that it is difficult to perform accurate astrometry at this time.  To determine the position of the central binary system requires high-resolution VLBI observations during the low/hard state, to detect the compact jet originating from the core of the system.  This requires the observations to be triggered sufficiently early in the outburst during the rise phase, or following the reverse transition back to the low/hard state at the end of the outburst, although the hysteresis effect \citep{Mac03} coupled with the radio/X-ray correlation in the low/hard state \citep*{Gal03} implies that the radio emission is less bright in the latter case.

\subsection{XTE J1752-223}

XTE J1752-223 was discovered on 2009 October 23 by the {\it Rossi X-ray Timing Explorer} ({\it RXTE}) during a routine scan of the Galactic bulge region \citep{Mar09a}.  The spectrum and lack of pulsations in the X-ray band led \citet{Mar09b} to suggest that the source was a black hole candidate, a conclusion supported by \citet{Mun10a} and \citet{Sha10}, who also claimed a black hole mass of 8--11\,$M_{\odot}$ and a source distance of $3.5\pm0.4$\,kpc from correlations between X-ray spectral and timing properties.  A 2-mJy radio counterpart was detected by \citet{Bro09} during the hard state of the system, before its proximity to the sun precluded detailed study by pointed instruments on board X-ray satellites.  As the source emerged from this zone of avoidance, a state transition from the hard X-ray spectral state to the intermediate state was detected \citep{Hom10,Sha10a,Neg10}.  The outburst was intensively monitored in the X-ray band by the {\it Monitor of All-sky X-ray Image (MAXI)} \citep{Nak10}, {\it RXTE} \citep{Sha10}, and {\it Swift} \citep{Cur11} satellites, which observed behaviour fairly typical for a black hole X-ray binary system.

The initial radio detection was followed up by further monitoring with the Australia Telescope Compact Array (ATCA), showing that the source had brightened by an order of magnitude by 2010 January 21.  A VLBI imaging campaign with the European VLBI Network (EVN) and Very Long Baseline Array (VLBA) detected moving, decelerating and expanding jet components \citep{Yan10}, suggesting an initially mildly relativistic jet.  While the radio core of the system was not detected, its location was inferred to be between the two components detected in the VLBA image of 2010 February 26.  Further target of opportunity VLBA observations were made in 2010 April following the return of the source to the hard X-ray spectral state.

The lack of an accurate core position rendered some of the interpretation of the VLBI images reliant upon various assumptions.  In this paper, we use optical astrometry in conjunction with VLBI radio imaging to determine the true core location, demonstrating how these two techniques can be highly complementary when used in parallel.  We describe our observations in Section~\ref{sec:observations}, present our results in Section~\ref{sec:results}, and re-interpret the existing VLBI observations in light of our newly-determined core position in Section~\ref{sec:discussion}.

\section{Observations and data reduction}
\label{sec:observations}

\subsection{Optical}

We observed the field containing XTE J1752-223 using the Inamori-Magellan Areal Camera and Spectrograph (IMACS) instrument mounted on the 6.5-m Baade Magellan telescope at Las Campanas Observatory. We obtained four $i^{\prime}$-band and four $g^{\prime}$-band images on 2009 November 3 00:10:26--00:19:28 UTC with exposure times of 5--10\,s.  The observing conditions were good with a photometric sky and a seeing of 0.8\arcsec\ and 0.9\arcsec\ in the $i^{\prime}$-band and $g^{\prime}$-bands, respectively.  Only a small section of one of the eight CCDs of the IMACS mosaic detector was read to sample a $\sim4^{\prime}\times 4^{\prime}$ field of view at 0.11\arcsec\,pixel$^{-1}$. 
 
\begin{figure*}
\centering
\includegraphics[width=\textwidth]{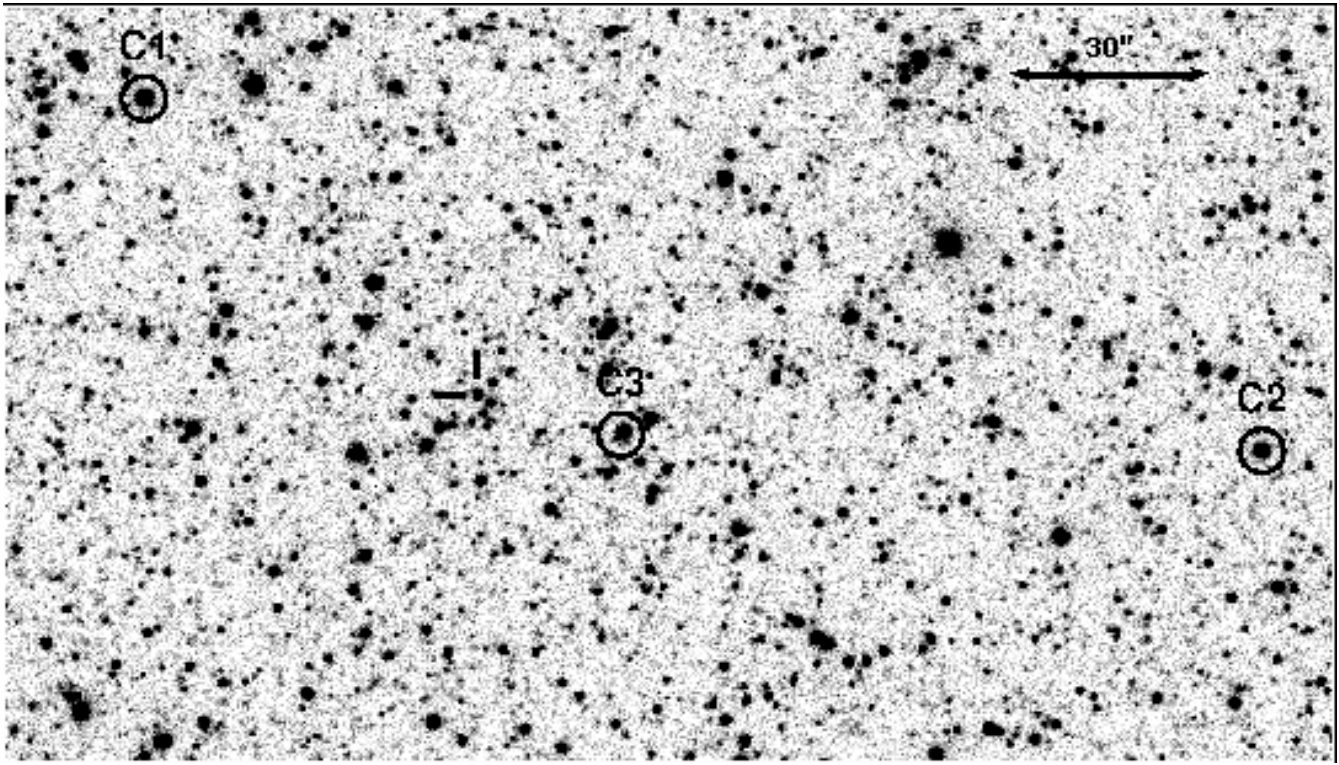}
\caption{Optical $i^{\prime}$-band finder chart for XTE J1752-223, taken from a 5-s image from 2009 November 3. North is up, east is to the left. The optical counterpart to XTE J1752-223 is indicated with the cross-hairs. The stars labeled C1 to C3 are comparison stars with $i^{\prime}$-band ($g^{\prime}$-band) magnitudes of 13.62 (14.66), 14.53 (16.25) and 14.38 (15.30), respectively. The accuracy in the above magnitudes is of order 1 per cent.}
\label{fig:finder}
\end{figure*}

We applied aperture photometry on each of the images using DAOPHOT within the Image Reduction and Analysis Facility ({\sc iraf}\footnote{\textsc{iraf} is distributed by the National Optical Astronomy Observatories}) software package to compute the instrumental magnitudes of the detected stars. Flux calibration of the field was performed by observing two Sloan Digital Sky Survey (SDSS) fields, and differential photometry was used to derive the source flux variability as a function of time. The photometric results given here are with respect to the three field stars C1--C3 shown in Fig.~\ref{fig:finder}.  The average $i^{\prime}$-band and $g^{\prime}$-band magnitudes of XTE J1752-223 were $16.22\pm0.05$ and $17.79\pm0.05$\,mag, respectively.

\subsection{VLBA}

At the end of its 2009--2010 outburst, XTE J1752-223 made a transition back to the hard state late in 2010 March \citep{Mun10b}.  As reported by \citet{Yan10}, two VLBA observations were made following this transition, on 2010 April 25 and 29 (MJD 55311.46 and 55315.46), under project code BB291, at an observing frequency of 5\,GHz.  The data were taken in dual-polarization mode with a recording rate of 512\,Mbps, corresponding to a total bandwidth of 64\,MHz per polarization.  XTE J1752-223 was phase referenced to the nearby calibrator source J1755-2232, from the third VLBA calibrator survey \citep[VCS3;][]{Pet05}.  The position assumed for the calibrator source was 17$^{\rm h}$55$^{\rm m}$26\fs285, $-22$\degr32\arcmin10\farcs593 (J\,2000), although the current best position\footnote{http://astrogeo.org/} is 17$^{\rm h}$55$^{\rm m}$26\fs2845, $-22$\degr32\arcmin10\farcs616 (J\,2000), with an uncertainty of 1.4\,mas.  Since phase referenced positions are measured relative to the position of the calibrator source, the measured target position must therefore be corrected by the 24-mas difference between the assumed and true positions.

In an attempt to perform accurate astrometry on the fading core of the system, one further epoch of VLBA observations was taken on 2010 June 17 (MJD 55364.31), at an observing frequency of 8.4\,GHz, under project code BM346.  The data were split into eight 8-MHz intermediate frequency pairs, corresponding to a total bandwidth of 64\,MHz in each of two independent polarizations.  We used the same phase reference source, J1755-2232, but assumed the best known position from the VLBA calibrator manual, as given above.  The phase referencing cycle time was 3\,min (2\,min on the target source, 1\,min on the calibrator), and we substituted every eighth scan on the target source for an observation of a nearby check source from the fifth VLBA calibrator survey \citep[VCS-5;][]{Kov07}, J1751-1950.  30\,min at the beginning and end of the observing run were used to observe bright calibrator sources at a wide range of elevations across the entire sky, in order to better calibrate unmodelled clock and tropospheric phase errors using the task DELZN from the Astronomical Image Processing System ({\sc aips}\footnote{{\sc aips} is produced and maintained by the National Radio Astronomy Observatory}) software package \citep{Gre03}, thereby improving the accuracy of the phase transfer and hence the resulting astrometry.

Data reduction was carried out according to standard procedures within {\sc aips}.  We corrected for small changes to the Earth orientation parameters used in the initial correlator model, and also for the ionospheric dispersive delay.  We used the measured system temperatures to calibrate the amplitude scale, and corrected the instrumental phase using the fringe finder source J1733-1304 from the international celestial reference frame \citep[ICRF;][]{Ma98}.  We performed fringe fitting on the phase reference source J1755-2232, which was then subjected to iterative imaging and self-calibration.  The final image was used as a model for bandpass calibration before transferring bandpass, amplitude and phase solutions to XTE J1752-223.  The strong scattering along the line of sight towards the source led us to use only the shortest baselines (up to 30\,M$\lambda(\nu/5{\rm GHz})$, where $\nu$ is the observing frequency) in making the images, tapering the weights of the visibility data with a Gaussian function of full width at half maximum 22.8\,M$\lambda$ to further downweight the long baselines.

\section{Results}
\label{sec:results}

\subsection{Optical}

We performed astrometry on a 5-second $i^{\prime}$-band exposure. We compared the positions of the stars against entries from the third U.S. Naval Observatory CCD Astrograph Catalog \citep[UCAC3;][]{Zac10}. An astrometric solution was computed by fitting for the reference point position, the scale and the position angle, considering all the sources that are not saturated and appear stellar and unblended. We obtain a solution with 0.031\arcsecond\ root-mean-square (rms) residuals from 26 stars (UCAC fit model magnitudes 14--16.5) well distributed on the $\sim$4\arcminute$\times$4\arcminute\ image. The positional accuracy of UCAC3 is estimated to be $\sim0.01$\arcsecond\ on stars of such magnitude \citep{Zac10}. In addition, the systematic uncertainty in tying the UCAC3 stars to the International Celestial Reference System (ICRS) is 5\,mas \citep{Zac10}.  For the accuracy on our stellar positions we adopt the linear sum of the residuals of the astrometry and the accuracy of the catalogue (as the latter is probably a systematic error): the resulting positional accuracy at 1$\sigma$ is 0.046\arcsecond\ on both right ascension and declination.  Thus our best optical position is
\begin{equation}
\begin{split}
\notag
{\rm RA} &= 17^{\rm h}52^{\rm m}15\fs093 \pm 0.003\\
{\rm Dec.} &= -22\degr20^{\prime}32\farcs35 \pm 0.05\qquad{\rm (J\,2000)},
\end{split}
\end{equation}
This is 533.9\,mas to the south east of the position of component A (corrected for the error in the assumed calibrator position) as identified by \citet{Yan10} in their VLBA observations of 2010 February 11.

\subsection{Radio}

\begin{figure}
\centering
\includegraphics[width=\columnwidth]{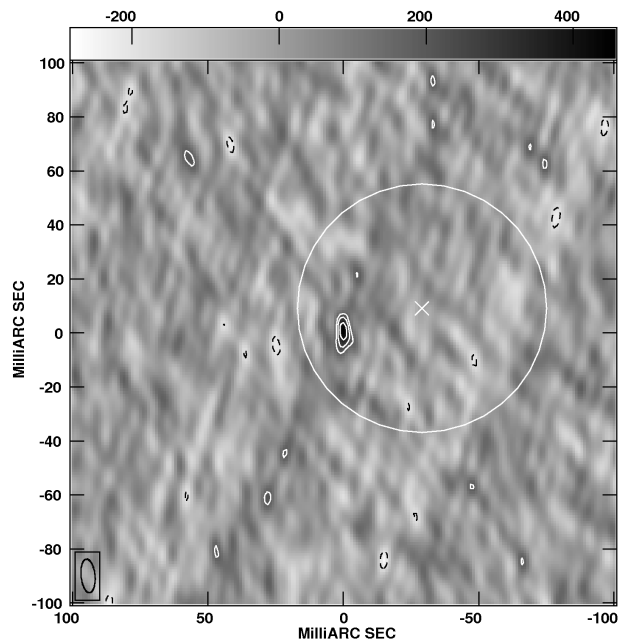}
\caption{5-GHz VLBA image of XTE J1752-223 on 2010 April 25.  Contour levels are at $\pm(\sqrt{2})^n$ times the rms noise of 0.066\,mJy\perbeam, where $n=-3,3,4,5...$  The optically-derived core position is marked with a cross, and the $1\sigma$ optical error circle of radius $0.046$\arcsecond\ is shown in white.  Co-ordinates have been corrected for the error in assumed calibrator position for the phase reference source.  The radio position lies well within the optical error circle, giving us confidence that we have correctly identified the radio core of the system.}
\label{fig:image}
\end{figure}

In the VLBA observations of 2010 April 25, a point-like radio source was detected with the VLBA, at a level of $0.46\pm0.07$\,mJy\perbeam, at a position consistent with the optical position of XTE J1752-223  (Fig.~\ref{fig:image}).  After correcting for the error in the assumed calibrator position, the position of the radio source, relative to the VCS-3 calibrator J1755-2232 (with an assumed position of (J\,2000) 17$^{\rm h}$55$^{\rm m}$26\fs2845, $-22$\degr32\arcmin10\farcs616), was
\begin{equation}
\begin{split}
\notag
{\rm RA} &= 17^{\rm h}52^{\rm m}15\fs09509 \pm 0.00002\\
{\rm Dec.} &= -22\degr20\arcmin32\farcs3591 \pm 0.0008\qquad{\rm (J\,2000)},
\end{split}
\end{equation}
where the quoted uncertainties are only the statistical errors from the Gaussian fitting algorithm (JMFIT within {\sc aips}).  The systematic error due to the cumulative effect of the error in the assumed calibrator position and the 0.76\degr\ throw between calibrator and target \citep*{Pra06} is estimated to be 0.51\,mas.  This VLBA position differs from the optically-derived core position by 30.4\,mas, along a position angle 107\degree\ E of N.  With the derived uncertainty of 46\,mas in both co-ordinates of the optical position, the radio source is well within the optical error circle.  Given the positional agreement, and since XTE J1752-223 was in the hard spectral state at the time of the radio observations, we conclude that this radio source is indeed a compact jet from the core of the system.  We note that the $\sim6$-month time offset between the observations from which the optical and radio positions were determined will likely lead to a small positional shift between epochs owing to the parallax and proper motion signatures of the source, of order 1\,mas if the source is at a distance of a few kpc and participates in the Galactic rotation.  Furthermore, since we see the radio emission from optical depth 1 at each frequency in a compact jet, the true core could be slightly offset along the jet axis from the position of the radio source.

On 2010 April 29, we detected a radio source at this same position, but it had faded to $0.29\pm0.08$\,mJy\perbeam\ at 5\,GHz.  On 2010 June 17, the source was only marginally detected, at a level of $0.25\pm0.08$\,mJy\perbeam\ at 8.4\,GHz.  The persistent, variable radio emission at this location also supports our conclusion that this radio source does indeed correspond to a compact jet from the core of the system.  In none of these three hard state observations was the radio core resolved.  Our best constraint on the size scale of the compact jet comes from the first, brightest observation on 2010 April 25, when the source was unresolved down to the beam size of $12.5\times5.2$\,mas$^2$ in position angle 5.5\degree\ east of north.  This corresponds to a source size of $<12.5(d/{\rm kpc})$\,au, where $d$ is the source distance \citep[i.e.\ $<44$\,au for the distance of 3.5\,kpc claimed by][]{Sha10}.

Following the detection of the radio core, we re-reduced the VLBA data of 2010 February 18, 23 and 26 (program code BB290) presented by \citet{Yan10}, using a larger image size to search for receding ejecta to the south-east of the core.  No new components were detected, to $5\sigma$ upper limits of 0.62, 0.74, and 0.45\,mJy\perbeam\ respectively.

\section{Discussion}
\label{sec:discussion}

Using high-precision optical astrometry, we have been able to locate the core of the X-ray binary system XTE J1752-223.  From VLBI observations made during the hard spectral state of the system, when the radio emission is dominated by a compact, unresolved core jet, we were able to further refine the core position.  In light of this new information, we now reanalyze the VLBI data presented by \citet{Yan10}.

\subsection{A double ejection event}
\label{sec:double}
The VLBA image of 2010 February 26 published by \citet{Yan10} shows two components, labelled A and B. While component A was detected in all four VLBI images from 2010 February, component B is detected in only this one image, and was interpreted by \citet{Yan10} as a receding component, leading the authors to infer that the true core of the system lay between components A and B in this image.  Our new determination of the true core position implies that both these components are located to the north-west of the core (Table~\ref{tab:components}).  From this, we infer that components A and B must arise from separate ejection events.  While we are unable to reliably pinpoint the exact epoch of ejection of each of those events, we can use X-ray spectral and timing information together with constraints from the integrated radio light curves to obtain a rough estimate for component A.

\begin{table}
\begin{center}
\scriptsize
\begin{tabular}{llccc}
\hline\hline
Component & Date & MJD & Separation & P.A.\\
& & (d) & (mas) & ($^{\circ}$)\\
\hline
A & 11-Feb-2010 & 55238.4 & $562.2\pm0.7$ & $-51.3\pm0.1$\\
A & 18-Feb-2010 & 55245.6 & $619.0\pm1.2$ & $-50.3\pm0.1$\\
A & 23-Feb-2010 & 55250.6 & $648.9\pm2.5$ & $-51.0\pm0.2$\\
A & 26-Feb-2010 & 55253.6 & $663.1\pm1.6$ & $-50.9\pm0.1$\\
B & 26-Feb-2010 & 55253.6 & $175.1\pm1.9$ & $-49.9\pm0.6$\\
\hline
\end{tabular}
\end{center}
{\caption{\label{tab:components}Angular separation from our newly-determined core position of the VLBI components detected by \citet{Yan10}.}}
\end{table}

\subsection{The ejection of component A}
\label{sec:ejection}
Radio flares in X-ray binaries have been linked to the transition from a hard intermediate state (HIMS) to a soft intermediate state (SIMS) during a rapid phase of X-ray spectral softening at the peak of the outburst \citep{Fen04}.  Also associated with this transition are a sharp drop in the integrated rms variability of the X-ray emission and a reduction in the coherence of the associated quasi-periodic oscillations (QPOs), from high-coherence Type C QPOs associated with flat-topped noise in the power spectrum, to lower-coherence Type A or Type B QPOs associated with weak red noise \citep{Bel05}.  However, we note that \citet{Fen09} found that while these changes in the variability properties were closely associated with radio ejection events, the association was not exact, such that one could precede the other by up to a few days.

On MJD\,55215.9, XTE J1752-223 was in a HIMS with a 2.2\,Hz Type C QPO \citep{Sha10}.  The integrated rms variability then decreased from 25 to 18 per cent as the QPO frequency rose to 5.3\,Hz by MJD\,55217.9, and by MJD\,55218.8, the observed QPOs had changed from Type C to Type A/B.  This suggests that the transition from HIMS to SIMS occurred around MJD\,55218.  Supporting this inference is the bright (20\,mJy), flat-spectrum radio emission observed on MJD\,55217 between 1.2 and 19\,GHz \citep{Bro10}.  This represents an increase of the radio brightness by an order of magnitude as compared to the initial 2-mJy radio detection in the rising hard state \citep{Bro09}, suggesting the onset of a radio flare.  This increase in radio brightness corresponds to an increase of only a factor of $\sim 2$ in the 15--50\,keV {\it Swift}/BAT and 4--10\,keV {\it MAXI}/GSC X-ray count rates \citep{Nak10} over the same period, which is not consistent with the radio/X-ray luminosity correlation \citep{Gal03} found for the compact jets of many hard state black hole candidates \citep[although note an ever-increasing number of outliers; e.g.][]{Gal07}.  While this would tend to support the interpretation of a radio flare on MJD\,55217, the flat radio spectrum over more than a decade in frequency instead argues that this radio emission most probably still arises from a compact jet, rather than discrete, optically-thin transient ejecta.  The anomalously bright radio emission could then correspond to the period of jet instability known to occur immediately preceding a large radio flare \citep{Fen04}.

Further radio information is available from the ATCA monitoring, which covered the entire outburst from the initial rising hard state through to the decay back to quiescence.  Although a full analysis is beyond the scope of this paper and will be presented by Brocksopp et al.\ (in prep.), we summarize the relevant information here.  The integrated radio light curve shows at least two large flares, followed by a few smaller events.  The last flat-spectrum radio detection was made on MJD\,55217, after which the 9-GHz radio flux density dropped to 3.3\,mJy on MJD\,55220, before peaking at 9.9\,mJy on MJD\,55221.  This suggests an initial ejection date between MJD\,55217 and 55220.  The second flare was somewhat broader, with the rise phase beginning after MJD\,55226 and the flare peaking at 10.9\,mJy at 9\,GHz on MJD\,55242.  The double-peaked light curve supports our conclusion from Section~\ref{sec:double} that the outburst comprised at least two ejection events.

\citet{Neg10} reported a sharp increase of the soft ($<4$\,keV) X-ray flux and a decline of the hard ($>10$\,keV) flux on MJD 55218, with the emergence of a disk blackbody component in the X-ray spectrum.  This is consistent with the evidence from the X-ray timing and radio observations, suggesting MJD\,55218 as the likely date of the initial radio ejection event.

\subsection{Deceleration of the ejecta}
\label{sec:deceleration}

\begin{figure}
\centering
\includegraphics[width=\columnwidth]{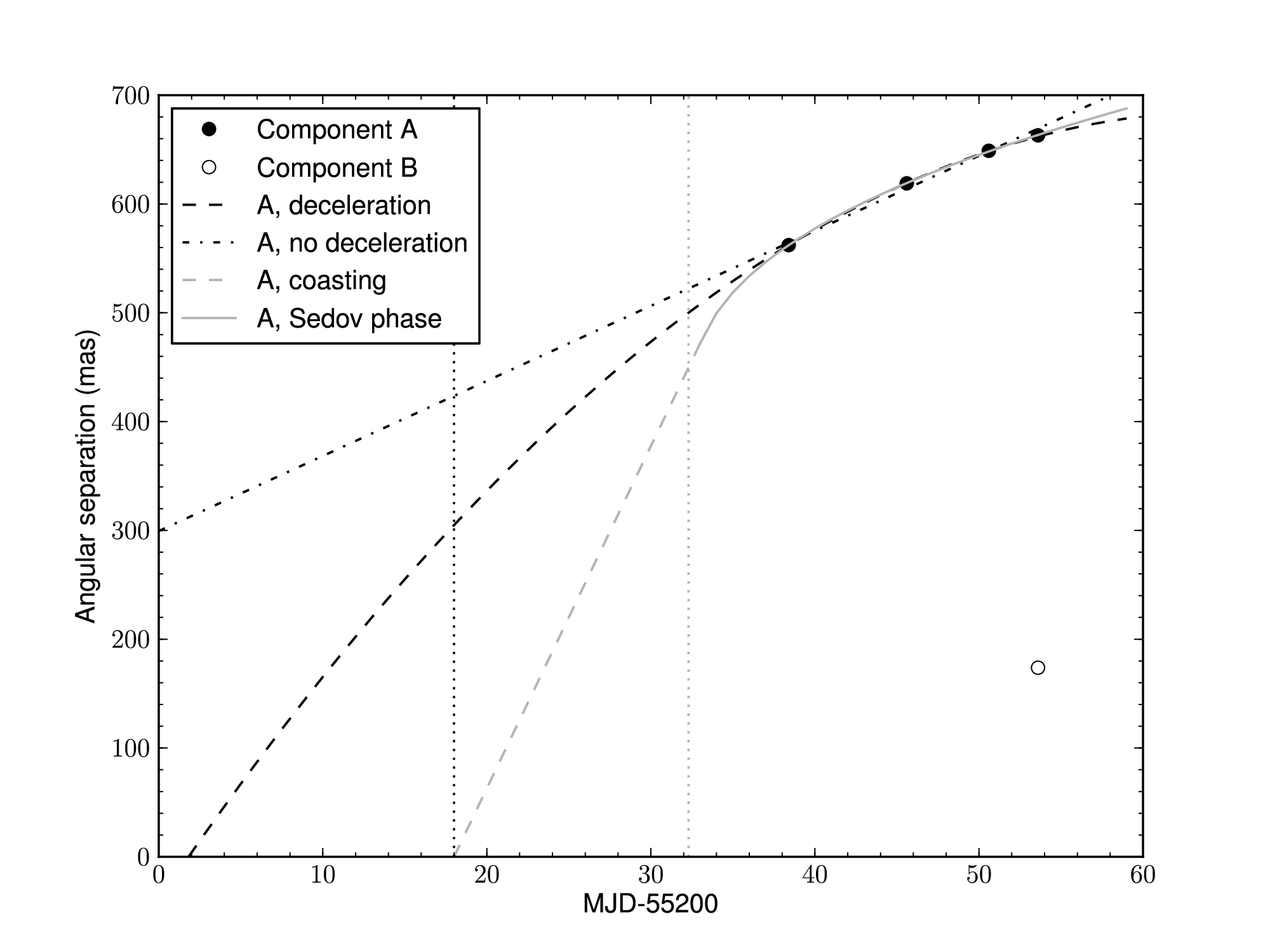}
\caption{Angular separation of the components detected by \citet{Yan10} from our newly-determined core position, together with their decelerating ejecta model (dashed line) and their ballistic ejecta model (dot-dashed line).  Filled points represent component A and the open point component B.  Error bars (0.7--2.5\,mas) are smaller than the marker size.  The black vertical dotted line indicates the time of the transition between HIMS and SIMS on MJD\,55218, as determined by \citet{Sha10}.  Grey line shows a possible model for the data, consisting of a period of pure ballistic motion lasting until $t_0$ (indicated by the grey vertical dotted line), followed by our fitted Sedov phase (Section~\ref{sec:deceleration}).  For the assumed ejection date, this appears to provide a better match to the data than either the pure ballistic model or the pure deceleration model.}
\label{fig:angsep}
\end{figure}

\citet{Yan10} found that uniform deceleration of component A fitted their measurements better than a ballistic model with no deceleration.  However, their suggested deceleration parameters (Fig.~\ref{fig:angsep}) imply an ejection date of MJD\,55201.8 (2010 January 5).  {\it RXTE} was still sun-constrained on this date, so no PCA observations are available to ascertain the X-ray state of the source, but the {\it Swift} and {\it MAXI} observations \citep{Nak10} show that this was significantly prior to the beginning of the X-ray spectral softening, and we deem this unlikely as the true date of ejection.

Assuming an ejection date of MJD\,55218, we are unable to fit the motion of the ejecta with a uniform deceleration model.  Our best-fitting model has a reduced-$\chi^2$ value of 230.9.  Either our assumed ejection date is wrong or the uniform deceleration model does not describe the data well.  A plausible alternative could be the scenario outlined by \citet*{Wan03}, whereby a shock wave propagates into the interstellar medium, sweeping up material as it moves, and decelerating such that the late-time behaviour approaches the Sedov solution, $R\propto t^{2/5}$.  Allowing the zero point to float and fitting the measured positions of component A with a simplistic Sedov model $R=R_0+k(t-t_0)^{0.4}$, where $R$ is the angular separation of the component from the core, we find $R_0=406\pm29$\,mas, $k=75.7\pm6.5$\,mas\,d$^{-0.4}$, and $t_0={\rm MJD\,}55232.3\pm1.6$, with a reduced $\chi^2$ value of 0.2.  Although the zero time and position ($t_0$ and $R_0$, respectively) do not correspond to our assumed ejection date and derived core position, when coupled with an initial coasting phase where the ejecta travel purely ballistically \citep[as proposed by][]{Hao09}, this model appears to provide a plausible fit to the data (Fig.~\ref{fig:angsep}).  

The derived zero time, $t_0$, is just consistent within error bars with an extrapolation of the expansion of component A \citep{Yan10} to zero size, which occurs on MJD\,$55229.7\pm1.0$.  It also coincides with the rise phase of the second flare in the integrated radio light curves, suggesting that the breadth of the second flare could be due to the release of energy as component A begins to decelerate, possibly combined with the ejection of component B.  However, if component B were ejected during this second radio flare, its non-detection in the three VLBI observations prior to MJD\,55253 is surprising. One explanation could be that the intrinsic jet speed and inclination angle to the line of sight are both high enough for the emission to be Doppler-deboosted until the jets have decelerated by sweeping up the surrounding gas.  Alternatively, if the observed jet ejecta are shocks, the delay in the appearance of component B could arise from the time taken for the ejecta to either catch up with the slower-moving material ahead of them (for internal shocks) or to sweep up and interact with the surrounding gas (for external shocks).  With only the one VLBI detection of component B and the lack of any signatures in the X-ray light curves that might correspond to a second ejection event, we cannot further constrain the ejection date of component B, and do not discuss it further.

In the absence of a precise ejection date for component A, constraints from the receding components, or from X-ray lightcurves of the ejecta as they decelerate, there are too few constraints to conduct a more meaningful fit to the full model of \citet{Hao09}.  However, the measured VLBI angular separations, the integrated radio light curves and the expansion of component A are all consistent with the deceleration, brightening and lateral expansion of that component close to MJD\,55232, as derived from our model fitting.  Thus, while we cannot definitively verify the proposed scenario, it is certainly plausible.  Should the model be applicable, the angular scale for deceleration ($<0.56$\arcsecond) would be significantly smaller for XTE J1752-223 than those derived by \citet{Hao09} for XTE J1550-564 (12--17\arcsecond) and H\,1743-322 (3\arcsecond).  While the distance is not yet well-determined, if XTE J1752-223 is indeed relatively nearby, as implied by the low hydrogen column towards the source \citep{Mar09b,Cur11} and as derived via a more model-dependent method \citep[$3.5\pm0.4$\,kpc;][]{Sha10}, then the discrepancy in the physical scale of the radius at which deceleration begins would be greater still.

\subsection{Quenching of the radio core in the soft state}

Knowing the true core position, we can constrain the quenching factor of the compact jet in the soft state.  During the three VLBA observations of 2010 February \citep{Yan10}, the core radio flux was in all cases $<2.4\sigma$, corresponding to a flux density of $<0.35$\,mJy\perbeam.  As compared to the 20\,mJy flux density measured on 2010 January 21, this represents a core quenching factor of $>57$.  This is consistent with previous lower limits on the quenching factor of the compact core jets in the soft X-ray state \citep[e.g.][]{Fen99}.

\subsection{No receding ejecta}

The reinterpretation of component B of \citet{Yan10} as an approaching component implies that their estimates of jet speed and inclination angle to the line of sight are no longer valid, since they relied on the ratios of sizes and flux densities of approaching and receding components.  Since a re-examination of the VLBA data from 2010 February did not show any receding components (Section~\ref{sec:results}), we cannot make revised estimates of the physical parameters of the jets with any accuracy.  However, assuming symmetric approaching and receding ejecta, linear expansion of the components \citep[as fitted for component A of XTE J1752-223 by][]{Yan10} and a value for the index of the electron energy spectrum, $p$, \citet*{Mil04} demonstrated that the ratio of the flux densities of approaching and receding components in a single image could be used to determine the product $\beta\cos\theta$.  Using the measured flux density of 2.2\,mJy for component A on 2010 February 18, and taking the $5\sigma$ upper limit of 0.62\,mJy for the flux density of the receding component, then assuming a canonical value for optically-thin ejecta of $p=2.2$, we find a lower limit $\beta\cos\theta>0.66$, such that $\beta\geq0.66$ and $\theta\leq49^{\circ}$.  As a caveat, we note that if significant deceleration and consequent brightening of the ejecta has occurred prior to this image being taken, as is conceivable (Fig.~\ref{fig:angsep} and Section~\ref{sec:deceleration}), the flux densities of approaching and receding components will no longer be governed by symmetric ejection, adiabatic expansion and Doppler boosting, so this analysis would not be valid.  Should the upper limit on $\theta$ be valid however, it would argue that the time delay hypothesis proposed in Section~\ref{sec:deceleration} is more probable than Doppler deboosting as an explanation for the delayed appearance of component B.

\subsection{Future proper motion studies}

The time baseline between the three radio observations in 2010 April and June is not sufficiently large to measure a significant positional shift between epochs, particularly given the low significance of the latter two detections.  Also, despite the 6-month time baseline between the observations from which the optical and radio positions were determined, the uncertainties are sufficiently large that we cannot reliably determine the source proper motion between these epochs.  However, our high-precision measurement of the core position provides an initial data point for a measurement of the proper motion should the source undergo repeated outbursts in the future, or should it be sufficiently bright in quiescence to be detected with the new generation of sensitive radio instruments such as the Expanded Very Large Array (EVLA) or the High Sensitivity Array (HSA) following the completion of the ongoing bandwidth upgrade at the VLBA.  Long time baselines and high-precision astrometry are essential for the measurement of X-ray binary proper motions, and given the relative rarity of outbursts in the majority of sources and the relatively small proper motion signals (typically a few milliarcseconds per year), it is important to take astrometric data at every possible opportunity.  We encourage future astrometric observations of this source should it return to a bright hard state.  A measurement of the source proper motion would provide information about the formation mechanism of the compact object \citep[e.g.][]{Mir01,Mir03}.

\section{Conclusions}

We have demonstrated how accurate optical astrometry can be used in conjunction with high-resolution VLBI imaging in the hard state of an X-ray binary to locate the compact core of the system to sub-milliarcsecond accuracy.  Our determination of the position of the core of XTE J1752-223 mandated a re-interpretation of the published VLBI data from the 2009--2010 outburst of the source.  The two components detected in the VLBA image of 2010 February 26 are both on the same side of the core, implying that there were at least two ejection events during the outburst, with the first likely occurring close to the transition from the hard intermediate state to the soft intermediate state on MJD\,55218.  With this extra constraint on the motion of component A, its angular separation from the core as a function of time can no longer be fit with a uniform deceleration model.  A plausible explanation could be ballistic motion out to some radius after which rapid deceleration occurred as the ejecta swept up the surrounding interstellar medium, and the jets transitioned to a Sedov phase.  We constrain the quenching factor of the compact core radio jet in the soft state to be $>57$.  No receding ejecta are detected in any of the VLBI observations, and from the upper limit to their flux density, we constrain the product of jet speed and inclination angle, $\beta\cos\theta$, to be $>0.66$.

\section*{Acknowledgments}

PGJ acknowledges support from the Netherlands Organisation for
Scientific Research via a VIDI grant.  The VLBA is operated by the
National Radio Astronomy Observatory, a facility of the National
Science Foundation operated under cooperative agreement by Associated
Universities, Inc.  This paper includes data gathered with the 6.5-m
Magellan Telescopes located at the Las Campanas Observatory,
Chile. This research has made use of NASA's Astrophysics Data System.

\label{lastpage}
\bibliographystyle{mn2e}

\end{document}